\documentstyle[pre,twocolumn,aps,epsf]{revtex}

\newcommand{\beq}{\begin{equation}}
\newcommand{\eeq}{\end{equation}}
\newcommand{\bea}{\begin{eqnarray}}
\newcommand{\eea}{\end{eqnarray}}
\newcommand{\req}[1]{Eq.~(\ref{#1})}
\newcommand{\gcc}{{\rm~g\,cm}^{-3}}
\newcommand{\dd}{{\rm\,d}}
\newcommand{\etal}{{\it\ et al.}}
\newcommand{\kp}{K_\perp}                      
\newcommand{\kpvec}{{\bf K}_\perp}
\newcommand{\NL}{N}
\newcommand{\mH}{m_{\rm H}}
\newcommand{\ApJ}[1]{Astrophys.~J. {\bf #1}}
\newcommand{\AandA}[1]{Astron.\ Astrophys. {\bf #1}}
\newcommand{\JPB}[1]{J.\ Phys.\ B {\bf #1}}
\newcommand{\PRL}[1]{Phys.\ Rev.\ Lett. {\bf #1}}
\newcommand{\PR}[2]{Phys.\ Rev.\ #1 {\bf #2}}

\begin{document}
\draft
\title{Partially ionized hydrogen plasma
in strong magnetic fields\thanks{{\it Phys.\ Rev. E\/} {\bf 60}, 2193--2208 (1999)}
}
\author{Alexander Y.~Potekhin,$^{1,}$\thanks{%
Electronic address: palex@astro.ioffe.rssi.ru}
 Gilles Chabrier,$^2$
 and Yuri A.~Shibanov$^1$}
\address{$^1$Ioffe Physical-Technical Institute,
     194021 St.\ Petersburg, Russia}
\address{$^2$%
CRAL (UMR CNRS No.\ 5574),
     Ecole Normale Sup\'erieure de Lyon,
     69364 Lyon Cedex 07, France}

\date{Received 10 February 1999}
\maketitle

\begin{abstract}
We study the thermodynamic properties 
of a partially ionized hydrogen plasma
in strong magnetic fields, $B\sim10^{12}-10^{13}$~G,
typical of neutron stars.
The properties of the plasma depend significantly
on the quantum-mechanical sizes and binding energies of 
the atoms, which are strongly modified by thermal motion
across the field. 
We use new fitting formulas for the atomic binding energies
and sizes, based on accurate numerical calculations
and valid for any state  of  motion of the atom.
In particular, we take into account
decentered atomic states, neglected in previous studies 
of thermodynamics of magnetized plasmas.
We also employ analytic fits for the thermodynamic functions of
nonideal fully ionized electron-ion Coulomb plasmas.
This enables us to construct an analytic model of the
free energy.
An ionization equilibrium equation is derived, 
taking into account the strong magnetic field
effects and the nonideality effects.
This equation is solved by an iteration technique.
Ionization degrees, occupancies, and the equation of state
are calculated.

\end{abstract}


\section{Introduction}
Magnetic fields $B\sim10^{12}-10^{13}$~G typical of 
isolated neutron stars qualitatively modify many 
physical properties of matter (see Refs.\cite{CanutoVentura,YaK}
for reviews).
In this paper we calculate the thermodynamic properties
of a strongly magnetized hydrogen plasma at 
temperature $T\sim10^{5.0}-10^{6.5}$~K,
which may compose outer neutron-star envelopes\cite{Pavlov,CPY,Page}.
As we shall see, the plasma under these conditions can be partially ionized,
and the quantum-mechanical properties of both free electrons
and bound species (primarily hydrogen atoms) are strongly modified by
the field, which thereby affects the thermodynamics.

The motion of charged particles in a magnetic field
is quantized into Landau orbitals.
The magnetic field is called {\it strongly quantizing\/}
if the free electrons populate mostly the ground Landau level\cite{YaK}.
This is the situation which we are especially interested in.
It occurs when the electron cyclotron energy 
$\hbar\omega_c=\hbar eB/(m_e c)$ 
(where $\hbar$, $e$, $m_e$ and $c$
are the Planck constant, electron charge, electron mass and 
speed of light, respectively)
exceeds both the thermal energy $k_B T$
and the electron Fermi energy $\epsilon_F$ --- that is for temperatures 
$T\ll T_B$ and densities $\rho<\rho_B$, where
\beq
   T_B=3.16\times 10^5\,\gamma{\rm~K} , 
\quad 
   \rho_B=0.809\,\gamma^{3/2}\gcc
\label{strong-field}
\eeq
(see Sec.~\ref{sect-electrons}).
Here, the parameter 
$\gamma=\hbar^3 B/(m_e^2 c e^3)=B/(2.35\times10^9$~G)
is the electron cyclotron energy in atomic units.

We will refer to a {\it strong\/} magnetic field
when $\gamma\gg1$.
A number of studies of the equation of state (EOS) of matter
in strong magnetic fields were based on various modifications
of the Thomas-Fermi approximation\cite{Const,Fushiki,AS91,Ossi}.
This approximation works reasonably well 
at large $\rho$ and for large ion charge $Z_i$.
Abrahams and Shapiro\cite{AS91} estimate
its validity range as $\rho\gg\rho_B Z_i^{-1/2}$.
We consider $Z_i=1$ and lower densities,
for which atoms are present in the plasma
and contribute to the EOS. 

The atom in a strong magnetic field $\gamma\gg1$ is compressed in 
the transverse directions to the size of the ``magnetic length'':
\beq
   a_m=(\hbar c/eB)^{1/2}=a_0\,\gamma^{-1/2},
\eeq
where $a_0=\hbar^2/(m_e e^2)$ is the Bohr radius.
The ground-state binding energy grows 
logarithmically with $B$
and exceeds the ground-state energy of the field-free atom 
by order of magnitude at $B\sim10^{12}$~G\cite{CanutoVentura}.
Ionization equilibrium of atoms
in strong magnetic fields was first discussed 
by Gnedin\etal\cite{Gnedin-etal}
and Khersonskii\cite{Kher-1}. 
Khersonskii\cite{Kher}
considered also dissociation equilibrium
of ${{\rm H}_2}^+$ species. 
However, these pioneering works neglected modifications 
of the atomic properties caused by the thermal motion
of the atoms across the field.

The motional modifications of quantum-mechanical characteristics 
of the atom arise from the coupling
between the center-of-mass motion  
across the field and the relative electron-proton
motion\cite{GD,Burkova,VDB,PM93,P94}.
The role of these effects was appreciated
by Ventura\etal\cite{Ventura-etal}, 
who, however, did not treat them quantitatively.
An increase of the nonionized fraction caused by
the motion effects was mentioned by Pavlov and M\'esz\'aros\cite{PM93},
who used perturbation theory applicable 
to atoms only slightly distorted from their rest-state
cylindrical shape. 
Quantum-mechanical calculations of 
binding energies and wave functions
of hydrogen atoms in {\em any\/} states of motion
in the strong magnetic
fields have been carried out only recently\cite{VDB,P94}.

Lai and Salpeter\cite{LS95,LS97} evaluated
the effects of motion on the ionization equilibrium
using an approximation for the 
binding energies of moving atoms which does not apply to
the so-called {\it decentered states},
for which the electron-proton separation is large
\cite{Burkova,VDB,P94}.
Nonideality effects were included in the ionization equilibrium
 equation only as a pressure-ionization factor 
for $\rho\gg10^2\gcc$.
As a result, this equation contains a 
factor which diverges (and becomes even negative) at sufficiently 
high temperatures. 

Recently, Steinberg\etal\cite{Steinberg}
calculated the second virial coefficient of the 
proton-electron plasma in arbitrary magnetic field 
and constructed an EOS at low densities.
The bound states were included 
using the Planck-Larkin partition function.
This approach yields correct EOS at the low density 
where the virial expansion holds\cite{EKK}.
However, the Planck-Larkin formalism fails 
at higher densities, where one has to resort to the chemical
picture of the plasma, as discussed in detail,
e.g., by D\"appen\etal\cite{Dappen87}.
In addition, atomic binding energies were calculated 
in \cite{Steinberg} using approximations \cite{LS95}
which have very restricted applicability
as shown in \cite{P98}.

In this paper we use new
fitting formulas to atomic energies and sizes\cite{P98}
based on a previous numerical study\cite{P94}, valid for 
any state of atomic motion.
The molecular H$_2$ fraction is evaluated following 
the approach of Lai and Salpeter\cite{LS97}
but with a modified treatment of nonideality.
Our knowledge of the quantum-mechanical properties of molecules
in a strong magnetic field is still incomplete, but the evaluation 
of the molecular fraction is useful 
to determine the validity domain of our EOS at relatively 
low temperatures (where the molecules dominate).

The next section presents a simple thermodynamic model
of the hydrogen plasma. The model is tested in the nonmagnetic case 
by comparison with more elaborate models,
and is shown to provide sufficient accuracy
at high $T$ where the molecular fraction is small.
In Sec.~\ref{sect-FI}, 
we consider a fully ionized plasma in a strong magnetic field.
The partial ionization and dissociation are discussed 
in Sec.~\ref{sect-PI}, where an analytic model
of the plasma free energy is constructed
and the ionization equilibrium equation
is derived. 
Numerical results are presented 
and discussed in Sec.~\ref{sect-res}.
\section{Thermodynamic model: the zero-field case}
\label{sect-thermo0}
\subsection{Chemical picture of the plasma}
\label{sect-chempict}
A theoretical description of partially ionized plasmas
can be based either on the physical picture or
on the chemical picture of the plasma\cite{EKK}.
In the chemical picture, 
bound species (atoms, etc.) are 
treated as elementary entities along 
with free electrons and nuclei. 
In the physical picture, nuclei and 
electrons (free and bound) are the only fundamental constituents of the 
thermodynamic ensemble. 
The relative merits of the two approaches have been 
discussed, e.g., in \cite{Dappen92,C94}.

We use the so-called occupation probability formalism
in frames of the chemical picture.
Occupation probabilities, which ensure convergence of
the internal partition functions (IPF),
were first introduced by Fermi\cite{Fermi},
who has demonstrated their immanent relation to 
a nonideal contribution in the Helmholtz free energy.
Various approaches to the construction of 
the occupation probabilities have been 
reviewed by Hummer and Mihalas\cite{HM}.
The approach adopted by 
Mihalas and co-workers \cite{HM,MDH,DMH} (hereafter MDH) 
is based on the Inglis-Teller
criterion of Stark broadening 
conventional for plasma spectroscopy, which gives
optical spectra consistent with available
experiments (see, e.g., Ref.\cite{Dappen87}).
However, the equation of state derived by MDH
is unrealistic at $\rho\gtrsim10^{-2}\gcc$ (see \cite{SCVH}),
and the approximations made in its derivation are
lacking in self-consistency \cite{P96}. An alternative EOS
was derived in a self-consistent manner
by Saumon and Chabrier\cite{C94,SCVH,SC91,SC} (hereafter SC)
from effective pair potentials between plasma particles,
but with neglect of the Stark broadening.
The ionization degree deduced by SC strongly differs
from that by MDH.
The origin of the discrepancy is rooted in the fact that 
strongly perturbed atoms, whose spectral lines disappear
due to the Stark merging,
may still contribute to the EOS as bound species
\cite{Rogers86}.
Thus the approaches of MDH and SC are reconciled
by an approximate treatment of 
the atoms perturbed by plasma ions 
as quasicontinuum atomic states,
which contribute to the EOS as atoms
although they do not show atomic spectral lines\cite{P96}.

The chemical picture faces a principal 
difficulty in cases where the interaction between
nuclei and electrons in a bound state is comparable to the
interaction between a bound object and neighboring plasma particles.
This situation occurs when
pressure ionization is important
or when high atomic levels are appreciably populated.
In these cases, a special term should be included into the
free-energy model, in order to distinguish 
between bound and free states.
For instance, MDH constructed an {\it ad hoc\/}
``pressure ionization term'' in the free energy \cite{MDH},
SC introduced hard cores with fixed diameters
in the effective potentials for bound species \cite{SC},
and exponential ``unbinding'' occupation
probabilities were used in \cite{P96}.
The latter approach has been justified
by considering an excluded volume of the bound objects
at relatively low density, assuming an uncorrelated
distribution of the plasma particles.
At high density, the strong correlations of
the positions of the particles must be taken into account.
Their approximate treatment in the hard-sphere model
(e.g., by SC) appears to be practical for this purpose.

In the case of the strong magnetic field,
the model of the plasma cannot be as detailed
as, e.g., the SC nonmagnetic model,
because the effective potentials
(partly derived from high-pressure experiments in the zero-field case)
are not available. Therefore we use a simple
hard-sphere picture described below.
In order to check the validity of this model, we apply
the same assumptions to the well-studied zero-field case
and compare the results with those of more elaborate models.

\subsection{Free-energy model}
\label{sect-zero-model}
Consider a plasma consisting of electrons, protons, and H atoms
in a volume $V$.
Let us write the Helmholtz free energy as
$
    F = F_{\rm id} + F_{\rm ex},
$
where 
\beq
   F_{\rm id}=F_{\rm id}^{(e)}+F_{\rm id}^{(p)}
           +F_{\rm id}^{\rm neu}+F_{\rm rad}
\eeq 
is the sum of the ideal-gas free energies of the electrons,
protons, neutral species, and photons (thermal radiation), respectively,
and $F_{\rm ex}$ is the {\it excess\/} (nonideal) part.

\subsubsection{Ideal part of the free energy}
We consider nondegenerate protons 
and neglect their spin statistics both in bound and free states
(this amounts to an additive constant in the entropy
that affects neither ionization equilibrium nor the EOS, 
provided the total number of 
free and bound protons $N_0$ is fixed).
Then
\beq
   \beta F_{\rm id}^{(p)} / N_p = 
        \ln(n_p\lambda_p^3) -1 ,
\eeq
where $\beta\equiv ( k_B T)^{-1}$.
Here and hereafter, $N_\alpha$, $n_\alpha$,
and $\lambda_\alpha\equiv (2\pi \beta\hbar^2/m_\alpha)^{1/2}$
denote, respectively, the total number, number density, and
thermal wavelength of particles of the type $\alpha$
with mass $m_\alpha$.

For the ideal gas of electrons,
we use the identity\cite{LaLi-stat}
\beq
   F_{\rm id}^{(e)}  = 
     \mu_e N_e - P_e V,
\label{Fe_id}
\eeq
where $\mu_e$ and $P_e$ are
the chemical potential and pressure of the ideal Fermi gas,
respectively, which can be obtained as functions of $n_e$ and $T$
from the equations
\bea
   P_e &=& {8\over3\sqrt\pi}\,{k_B T \over \lambda_e^3}\,
   I_{3/2}(\beta\mu_e),
\label{P_e0}
\\
   n_e &=& {4 \over \sqrt\pi\lambda_e^3}\,
   I_{1/2}(\beta\mu_e).
\label{n_e0}
\eea
Here, 
\beq
   I_\nu(z) \equiv \int_0^\infty
  { x^\nu \over \exp(x-z)+1 } \,\dd x 
\label{I_nu}
\eeq
is the Fermi integral. 
With the use of Pad\'e approximants to the functions $I_\nu(z)$
and their inverse functions\cite{Antia},
$F_{\rm id}^{(e)}$ is expressed
as an analytic function of $N_e$, $V$, and $T$.

In the zero-temperature limit,
one may replace $I_\nu(\beta\mu_e)$ 
by $(\beta \epsilon_F)^{\nu+1}/(\nu+1)$, which gives, in particular,
the well-known expression 
\beq
   \epsilon_F = {\hbar^2\over2m_e}\, (3\pi^2 n_e)^{2/3}.
\label{EF0}
\eeq
The Fermi temperature is defined as 
$T_F\equiv \epsilon_F/k_B\approx 3\times10^5\,\hat\rho^{2/3}$~K,
where $\hat\rho=1.6735 \, n_e/(10^{24}{\rm~cm}^{-3})$ 
is the mass density of the electron-proton plasma in $\gcc$.
In the nondegenerate limit $T \gg T_F$, the ideal Boltzmann gas
relations are recovered,
$\mu_e = k_B T \ln(n_e \lambda_e^3/2)$ and
$P_e = n_e k_B T$.

For the atoms, one has
\beq
    F_{\rm id}^H = k_B T \sum_\kappa N_\kappa  
    \left[\ln(n_\kappa\lambda_{\rm H}^3/g_\kappa) - 1 - 
    \beta\chi_\kappa \right],
\label{Fatom-id}
\eeq
where $\kappa$ enumerates quantum states with statistical weights 
$g_\kappa$ and binding energies $\chi_\kappa$.

It should be noted that, although nonideality effects are 
not included in $F_{\rm id}$ explicitly, they do affect
the equilibrium value of $F_{\rm id}$
through the particle numbers.
In particular, the distribution of $N_\kappa$ in \req{Fatom-id} is 
{\em not\/} assumed to obey the ideal-gas Boltzmann law.

Finally, the radiation term (which can be important only at 
low $\rho$ or very high $T$)
reads
\beq
   F_{\rm rad} = - (4\sigma/3c) \, V T^4,
\label{Frad}
\eeq
where $\sigma=\pi^2 k_B^4 / (60\hbar^3 c^2)$
is the Stefan-Boltzmann constant.

\subsubsection{Excess free energy}
The excess free energy is conventionally written as
\beq
   F_{\rm ex} = F_{\rm ex}^{C}+F_{\rm ex}^{\rm neu},
\label{Fex}
\eeq
where $F_{\rm ex}^{C}$
is the excess free energy of the ionized part of the plasma
and $F_{\rm ex}^{\rm neu}$ accounts for 
interactions of neutral species with electrons, protons, 
and other neutral species.
The Coulomb term 
\beq
   F_{\rm ex}^{C} = F_{ii}+F_{ee}+F_{ie}
\eeq
includes contributions from
the exchange and correlation interactions of electrons $F_{ee}$,
Coulomb interactions in the one-component plasma (OCP) of ions
$F_{ii}$, and ion-electron (screening) interaction $F_{ie}$.
These contributions have been calculated by various procedures,
e.g., by solving a set of hypernetted-chain equations
or Monte Carlo simulations\cite{BausHansen,IIT,C90,CP98,DWSC}.
We make use of the fitting formulas to the results
of such calculations, obtained in \cite{IIT}
for $F_{ee}$ and in \cite{CP98} for $F_{ii}$ and $F_{ie}$.
These formulas express the electron-ion plasma free energy
as an analytic function of the electron density parameter
\beq
   r_s = a_e/a_0\approx 1.39 \hat\rho^{-1/3}
\label{rs}
\eeq
and Coulomb coupling parameter
\beq
   \Gamma = \beta e^2/a_e 
    \approx 0.227\,\hat\rho^{1/3} / T_6,
\label{Gamma}
\eeq
where $a_e=(4\pi n_e/3)^{-1/3}$ is the mean interelectron distance
and $T_6\equiv T/10^6{\rm~K}$.

The nonideal part of the atomic free energy, $F_{\rm ex}^{\rm neu}$, 
can be written as\cite{SC91,SC}
\beq
    F_{\rm ex}^{\rm neu}= F_{\rm HS} + F_{\rm pert},
\eeq
where $F_{\rm HS}$ is the reference free energy, 
treated in the hard-sphere approximation, and 
$F_{\rm pert}$ is the perturbation part that
accounts for the attractive 
(van der Waals) interactions. To calculate these contributions,
an elaborate model has been developed by SC\cite{SC91,SC}.
Its simplified analytic version for weak electron degeneracy
has been constructed in \cite{P96}.
In the so-called van der Waals one-fluid model\cite{Jung-},
a free energy of the hard-sphere mixture is represented by
the Carnahan-Starling formula\cite{CarnahanStarling}
\beq
    \beta F_{\rm HS}/N_{\rm tot} = (4\eta-3\eta^2)/(1-\eta)^2,
\label{F_HS}
\eeq
where $N_{\rm tot}=\sum_\alpha N_\alpha$ 
is the total number of particles,
\beq
    \eta = {\pi\over 6 N_{\rm tot}V}\sum_{\alpha\alpha'} N_\alpha 
    N_{\alpha'}d^3_{\alpha\alpha'}
\label{eta}
\eeq
is the effective packing fraction, and $d_{\alpha\alpha'}$ are the
hard-sphere interaction diameters.
In our case, the subscript $\alpha$ enumerates atomic quantum
states described by quantum numbers
$\kappa$ and takes on a single value $p$ 
for the free protons.

In the following, we compare two versions 
of the model: ({i}) the {\it full\/} version,
in which $F_{\rm pert}$ and $d_{\alpha\alpha'}$
are given by approximations of \cite{P96},
with one exception (adopted from \cite{SC})
that $d_{\alpha\alpha'}$ cannot be smaller than
a certain limit $d_{\alpha\alpha'}^{(0)}$,
and ({ii}) the {\it simple\/} version,
in which long-range atomic interactions are disregarded.
In the latter case, $F_{\rm pert}=0$
and $d_{\alpha\alpha'} = d_{\alpha\alpha'}^{(0)}$
Furthermore, we adopt the simplest choice
\beq
   d_{\kappa\kappa'}^{(0)}=l_\kappa+l_{\kappa'} ,
\quad
  d_{\kappa p}^{(0)}=l_\kappa,
\label{diam}
\eeq
where $l_\kappa$ is the root-mean-square
proton-electron distance in the quantum state $\kappa$\cite{size}.
For the interactions among charged particles,
we define $d_{\alpha\alpha'}=0$, because this type of interaction
is already included in the Coulomb part of the free energy.
Note that in the second (simple) version of the model,
$F_{\rm HS}$ turns into
the unbinding term $F_{ub}$ of Ref.\cite{P96} in the low-density limit
($\eta\ll1$). Thus the unbinding term 
is now incorporated in $F_{\rm HS}$, which allows us to approximately
take into account the correlation effects.
\subsection{Equilibrium conditions}
\label{sect-Saha0}
Thermodynamic equilibrium is
given by the minimization of $F(V,T,\{N_\alpha\})$
with respect to the particle numbers $N_\alpha$ 
under stoichiometric constraints.
The condition of the extremum of $F$ can be written in the form
of the Saha equation 
corrected for nonideality and electron degeneracy:
\beq
   n_{\rm H} = n_p n_e \lambda_e^3 (m_p/m_{\rm H})^{3/2} 
          (Z_w/2) e^{\Lambda},
\label{Saha0}
\eeq
where
\beq
   \Lambda =
        \beta \partial F_{\rm id}^{(e)} / \partial N_e
     - \ln(n_e\lambda_e^3/2)
\eeq
allows for electron degeneracy,
and
\beq
    Z_w = \sum_\kappa g_\kappa w_\kappa {\rm e}^{\beta\chi_\kappa}
\label{IPF}
\eeq
is the modified IPF which includes the occupation probabilities $w_\kappa$,
defined according to\cite{P96}:
\beq
    k_{\rm B} T \ln w_\kappa = {\partial F_{\rm ex} \over 
    \partial N_p} + {\partial F_{\rm ex} \over \partial 
    N_e} - {\partial F_{\rm ex} \over \partial N_\kappa}. 
\label{occprob0}
\eeq
To solve \req{Saha0},
one must add the electroneutrality condition $n_e=n_p$
and the mass conservation condition
$n_{\rm H}+n_p=n_0$, where 
$n_0=\rho/m_{\rm H}=(\rho/11.293\gcc)\,a_0^{-3}$.

The Boltzmann distribution of the atoms, 
corrected for nonideality, reads
\beq
   n_\kappa = n_{\rm H}\, g_\kappa\, w_\kappa\, e^{\beta\chi_\kappa}/Z_w.
\label{Boltz}
\eeq

The minimum of the free energy is sought by solving
Eqs.~(\ref{Saha0})--(\ref{Boltz}) iteratively\cite{P96}.
First, one defines starting $w_\kappa$'s and calculates 
the number densities from 
Eqs.~(\ref{Saha0}) and (\ref{Boltz}). Then the
$w_\kappa$'s are refined using these number densities in 
\req{occprob0}\cite{endnote1}.

The molecules H$_2$ can be easily included in this procedure.
The dissociation--recombination equation reads 
\begin{equation}
    n_{{\rm H}_2} = n_{\rm H}^2 (\lambda_{\rm H} \sqrt{2})^3 
    Z_{w2} / Z_w^2, 
\label{disseq}
\end{equation}
where $Z_{w2}$ 
is the molecular IPF, modified 
by multiplying each $\kappa$th term by an occupation probability
$w_\kappa^{{\rm H}_2}$\cite{P96}, given by
\beq
    k_{\rm B} T \ln w_\kappa^{{\rm H}_2} = 2\left( 
    {\partial F_{\rm ex} \over 
    \partial N_p} + {\partial F_{\rm ex} \over \partial 
    N_e} \right)
    - {\partial F_{\rm ex} \over \partial N_\kappa^{{\rm H}_2}}.
\label{wH2}
\eeq
For simplicity, we do not include molecules in the present
versions of the model, because the fraction of H$_2$
is small in the range of $\rho$ and $T$ which we are interested in.
   
After the equilibrium distribution of plasma particles
is found, the pressure $P$, internal energy $U$,
and entropy $S$ are
calculated from the relations
\beq
   P=-(\partial F/\partial V)_{T,\{N_\alpha\}}, \quad
   U=[\partial(\beta F)/\partial\beta]_{V,\{N_\alpha\}},
\label{TDrel}
\eeq
$S=(U-F)/T$.
The higher-order thermodynamic 
quantities are obtained by differentiation
of $P,U,S$ without keeping $N_\alpha$ fixed\cite{LaLi-stat}.
\subsection{Results of comparison}
\label{sect-res0}
\begin{figure}
\epsfysize=55mm
\epsffile[68 260 400 565]{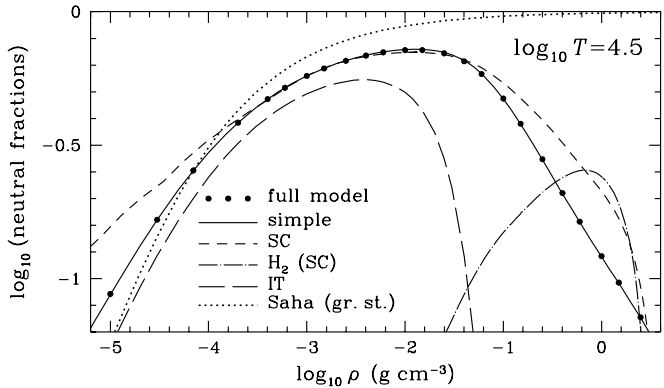} 
\caption{Comparison of the fraction of neutral atoms
$f_{\rm H}=n_{\rm H}/n_0$ given by SC tables (short-dashed line) and by
two present versions of the thermodynamic model 
of partially ionized atomic hydrogen (see text).
The long-dashed line corresponds to the fraction of atoms that 
satisfy the Inglis-Teller criterion.
The dotted line is given by the usual Saha equation
for ground-state atoms.
}
\label{fig-ieh}
\end{figure}

The ionization curves given by different versions
of the model are compared in Fig.~\ref{fig-ieh}
for $T=10^{4.5}$~K. 
Although the neglect of the perturbation terms
introduced in the simple version is most perceptible
at such relatively low temperatures, 
the ``full'' and ``simple'' versions of the model
yield practically identical atomic fractions $f_{\rm H}\equiv n_{\rm H}/n_0$.

The results of SC\cite{SCVH} qualitatively
agree with the present model. Quantitatively,
they differ in the pressure-ionization
region at $\rho>0.1\gcc$, where the theoretical 
uncertainty is largest (see Sec.~\ref{sect-chempict}).
The difference in the low-density regime $\rho< 10^{-4}\gcc$
is due to highly excited states.
If both the effective
diameter $d$ and statistical weight $g_n$ are
proportional to $n^2$ ($n$ being the principal quantum
number), then the neutral fraction should asymptotically 
decrease at low density
as $f_{\rm H}\propto\rho^{1/2}$. 
Our present model exhibits this asymptotic behavior;
the dependence $f_{\rm H}\propto\rho^{1/3}$ seen at low $\rho$
in the SC data might result from a choice $d\propto n$.

The long-dashed curve represents the fraction of
atoms satisfying the Inglis-Teller criterion:
 $f_{\rm IT}=\Sigma n_\kappa \tilde{w}_\kappa / n_0$.
Here, $\tilde{w}_\kappa$ is the probability that a given atom
is not strongly perturbed by plasma microfields;
it is estimated from Eq.~(31) of Ref.~\cite{P96}.
Using $f_{\rm IT}$,
we have calculated monochromatic opacities of the plasma
and compared them with the OPAL 
monochromatic opacities \cite{OPALo}
(at $\rho\leq10\,T_6^3\gcc$ where the OPAL data 
exist).
Along the isotherm shown in Fig.~\ref{fig-ieh},
our results agree with OPAL within 12\% 
in the photon energy range from 13.6 eV to 500 eV
where the opacity is dominated by bound-free atomic absorption. 
For comparison, hydrogen opacities calculated in \cite{ZPS}
differ from OPAL by up to $37\%$
(in the same range of energy and density
at the same $T$).

\begin{figure}
\epsfysize=90mm
\epsffile[70 160 560 640]{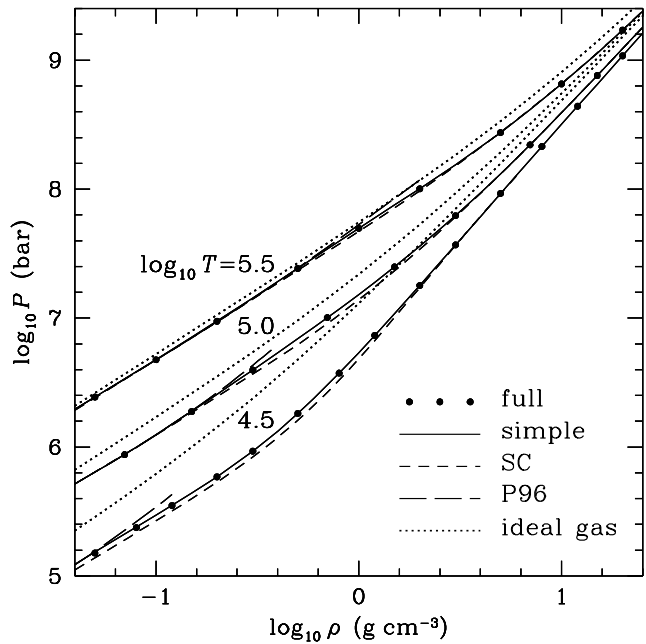} 
\caption{Comparison of two present versions of the EOS 
of partially ionized atomic hydrogen (see text)
with Refs.\protect\cite{SCVH} (SC)
and \protect\cite{P96} (P96).
The EOS of ideal fully ionized gas is also shown.
}
\label{fig-prh}
\end{figure}

Figure~\ref{fig-prh} demonstrates that the EOS
obtained with the full and simple versions of our model
practically coincide.
In the region of weak degeneracy, they also coincide
with the model presented in\cite{P96}.
Moreover, there is a good agreement with the SC model\cite{SCVH}.
Small differences occur only in the regions where
the SC model predicts an appreciable amount of molecules,
as explained in \cite{P96}.

\begin{figure}
\epsfysize=88mm
\epsffile[50 150 520 650]{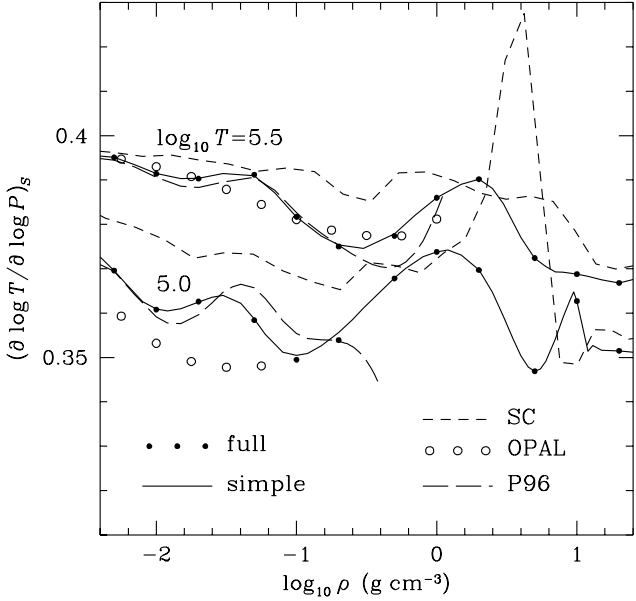} 
\caption{Two adiabatic gradient isotherms given by 
different EOS models 
in the domain of partially ionized atomic hydrogen.
}
\label{fig-tgh}
\end{figure}

As is well known, the second-order quantities
are more sensitive to the details of the thermodynamic model
than the first-order ones. The adiabatic temperature gradient
\beq
   \nabla_{\rm ad} = (\partial \ln T / \partial \ln P)_S
\eeq
is shown in Fig.~\ref{fig-tgh}.
There are only tiny differences between the full and simple
versions. For comparison, we also show $\nabla_{\rm ad}$
from other models. In its validity region (i.e., at low
density), the model\cite{P96} approximately agrees 
with the present one. The differences with 
predictions of SC are somewhat larger.
In all models, the isotherms ``wiggle'' in the region of 
consecutive pressure destruction
of excited atomic states. Such wiggles are absent
in the OPAL data\cite{OPAL}, based on the physical
picture of the plasma and also shown in Fig.~\ref{fig-tgh}.
Compared to SC,
the present data tend to be closer to the OPAL data.
We conclude that the simplifications
introduced above are acceptable to describe
the thermodynamics of atomic hydrogen.
In Sec.~\ref{sect-PI}, we generalize the 
model to the case of the strong magnetic field. 

\section{Fully ionized plasma in a strong magnetic field}
\label{sect-FI}
In this section, we describe effects of quantizing magnetic
fields on the fully ionized proton-electron plasma.
We assume that the field ${\bf B}$ is uniform and directed
along the $z$-axis.
\subsection{Ideal gas}
\label{sect-id}
\subsubsection{Electrons}
\label{sect-electrons}
The electron energy in a magnetic field reads\cite{LaLi-QM}
\beq
   \epsilon_\NL(p_z) = \NL\hbar\omega_c+p_z^2/(2m_e),
\label{magnenergy}
\eeq
where $p_z$ is the longitudinal momentum and
$\NL=0,1,2\ldots$ is the Landau quantum number.
All levels except the lowest one are double degenerate
with respect to the spin projection.
Strictly speaking,
the anomalous magnetic moment of an electron
leads to a splitting of the levels $\NL\geq1$ by
$0.00116\, \hbar\omega_c$,
which takes off the double degeneracy. 
However, 
this splitting cannot affect the thermodynamics at $\rho < \rho_B$,
where $k_B T$ should be at least comparable to $\hbar\omega_c$
for an appreciable population of the excited Landau levels.

The thermodynamic functions of the electron gas in the magnetic field
are easily derived from the first principles\cite{LaLi-stat}.
Taking into account the fact 
that the number of quantum states of an electron
with fixed discrete quantum numbers
in volume $V$ per longitudinal momentum interval $\Delta p_z$
equals $V\Delta p_z/(4\pi^2 a_m^2 \hbar)$\cite{LaLi-QM},
the thermodynamic potential $\Omega = -PV$ can be written as
\[
   \Omega = 
     -{V k_B T\over 2\pi^2 a_m^2 \hbar} \sum_{\NL=0}^\infty 
       g_\NL \int_0^\infty 
      \ln\left( 1+{\rm e}^{\beta[\mu_e-\epsilon_\NL(p_z)]}\right) \dd p_z,
\]
where
$g_{\NL}$  is the statistical weight
($g_0=1$ and $g_{\NL}=2$ for $\NL\geq1$).
Hence the electron pressure and number density are 
given by the equations
\bea
   P_e & = & {k_B T\over\pi^{3/2} a_m^2\lambda_e}
     \sum_{\NL=0}^\infty g_{\NL} I_{1/2}(\beta\mu_\NL),
\label{P_e}
\\
   n_e & = & {1\over2\pi^{3/2} a_m^2 \lambda_e}
     \sum_{\NL=0}^\infty g_{\NL} I_{-1/2}(\beta\mu_\NL),
\label{n_e}
\eea
where $\mu_\NL\equiv\mu_e-\NL\hbar\omega_c$.
The Helmholtz free energy is given by \req{Fe_id},
where $\mu_e$ can  be found by inversion of \req{n_e}
(e.g., using an algorithm described in \cite{PY}).

In the nonquantizing magnetic field $T_B\ll T$,
where many Landau levels are populated, the sum over $\NL$
in Eqs.~(\ref{P_e}), (\ref{n_e}) 
may be approximated by an integral, and integration by parts
 reproduces Eqs.~(\ref{P_e0}), (\ref{n_e0}).

In the domain of strong magnetic quantization, $T \ll T_B$ and
$\rho < \rho_B$, one may retain only the term $\NL=0$.
In that case, replacing $I_{-1/2}(\beta\mu_e)$ by 
$2\sqrt{\beta \epsilon_F}$ in \req{n_e} 
(by analogy with Sec.~\ref{sect-zero-model})
yields the Fermi energy
\beq
   \epsilon_F = {2\pi^4 \hbar^2\over m_e}\,(a_m^2 n_e)^2.
\label{EF}
\eeq
By definition, $\rho=\rho_B$ at $\epsilon_F = \hbar\omega_c$.
Hence $\rho/\rho_B=3\pi (2\gamma r_s^2)^{-3/2}$,
from which \req{strong-field} follows.
A comparison of Eqs.~(\ref{EF0}) and (\ref{EF})
reveals that the Fermi energy changes
by a factor $(4/3)^{2/3}(\rho/\rho_B)^{4/3}$.
Thus the degeneracy is strongly reduced at $\rho\ll\rho_B$.

In the nondegenerate regime $T\gg T_F$,
one has $I_\nu(\beta\mu_e)\approx\exp(\beta\mu_e)\Gamma(\nu+1)$;
therefore Eqs.~(\ref{P_e}), (\ref{n_e}) reduce to $P_e = n_e k_B T$ 
and 
\beq
   \beta\mu_e = \ln(n_e \lambda_e^3/2) - \ln u + \ln(\tanh u),
\label{mu_e}
\eeq
where $u\equiv\beta\hbar\omega_c/2=T_B/(2T)$.
This yields an explicit analytic form for $F_{\rm id}^{(e)}$.
In the nonquantizing field, $T_B\ll T$, the last two terms in
\req{mu_e} cancel out, and the classical 
expression (Sec.~\ref{sect-zero-model}) is recovered.
In the strongly quantizing regime,
$\rho<\rho_B$ and $T_F\ll T\ll T_B$, 
the last term of \req{mu_e} vanishes, 
which yields
\beq
   F_{\rm id}^{(e)} = 
      N_e k_B T \left[ \ln(2\pi a_m^2\lambda_e n_e) - 1 \right].
\label{Fe-magn}
\eeq
\subsubsection{Protons}
\label{sect-protons}
The transverse motion of the protons is quantized in Landau
orbitals with the elementary excitation equal to the
proton cyclotron energy $\hbar\omega_{cp}= (m_e/m_p)\hbar\omega_c$.
The energy spectrum is given by \req{magnenergy}
when replacing $m_e$ by $m_p$ and $\omega_c$ by $\omega_{cp}$.
Unlike the case of electrons, 
the double spin degeneracy of the Landau levels 
is taken off by the abnormal magnetic moment of the proton.

In our analysis, the protons are always nondegenerate,
so that by analogy with \req{Fe-magn} we have
\beq
   \beta F_{\rm id}^{(p)}/N_p =
      \ln(2\pi a_m^2\lambda_pn_p)
    + \ln\left[1-{\rm e}^{-\beta\hbar\omega_{cp}}\right]-1.
\label{Fp}
\eeq
Here, for sake of brevity, we drop the zero-point
energy $\case12\hbar\omega_{cp}$ and the spin energy
$\pm \case14 g_p\hbar\omega_{cp}$, where $g_p=5.585$
is the proton spin gyromagnetic factor\cite{LaLi-QM}.
We suppress these terms also for atoms and molecules.
Taking them into account yields an additive contribution 
to the total free energy of the system, equal to
\beq
  \Delta F = N_0 \left\{ \case12 \hbar\omega_{cp}
    - k_B T \ln[2\cosh(\beta g_p\hbar\omega_{cp}/4)] \right\}
\label{DeltaF}
\eeq
($N_0=N_p$ in the case of full ionization).
Since $N_0$ is constant, 
$\Delta F$ affects neither
ionization equilibrium nor pressure,
but it does affect the internal energy and specific heat,
therefore we take it into account in Sec.~\ref{sect-res}.
\subsection{Nonideal Coulomb plasma}
\label{sect-Coul}
According to the Bohr--van Leeuwen theorem,
magnetic field does not affect the thermodynamics 
of classical charged particle systems (see, e.g., Ref.\cite{Cornu}).
Thus the classical ionic OCP 
excess free energy $F_{ii}(\Gamma)$
does not depend on $B$ at any $\Gamma$.
The classical regime for the electron-proton plasma
corresponds to $r_s\gg1$ and $\Gamma\ll1$, where
the excess Coulomb free energy is given by
the Debye-H\"uckel formula
$
    F_{\rm ex}^{C} = - N_e k_B T \,\sqrt{8\Gamma^3/3}.
$
Indeed, it is easy to check that this law holds
independent of $B$ \cite{AS91}.

A magnetic field, however, affects
quantum-mechanical contributions to $F_{\rm ex}^{C}$.
These effects have been studied 
only in low-temperature or low-density regimes.

The ground-state exchange energy of the electron gas
in the strongly quantizing field \cite{Fushiki,Danz} behaves as
$- 2.25(\gamma r_s^3)^{-1}[\ln(\gamma r_s^2)-0.457+\cdots]\,e^2/a_0$
(per electron),
compared to $-0.75\pi^{-1}(9\pi/4)^{1/3}r_s^{-1}e^2/a_0$
in the nonmagnetic case\cite{Perrot}.
Thus the exchange energy at $T\ll T_F$ 
is suppressed by a factor 
$\approx 
          0.2036\,\gamma r_s^2/[\ln(\gamma r_s^2)-0.457]$.
Note that the condition 
of strong magnetic quantization requires $\gamma r_s^2$ 
to be large.

A general
low-density expansion for the free energy of a Coulomb plasma
in arbitrary magnetic field
up to order $\rho^{5/2}$ has been derived by Cornu \cite{Cornu}.
The coefficients of this expansion are not available
in explicit analytic form but require numerical evaluation,
which has not been done yet.
In the particular case of the OCP, 
a Wigner-Kirkwood-type expansion in powers 
of $\hbar$ is available \cite{Cornu}.
The lowest-order term of the latter expansion
(quantum diffraction term
of order $\hbar^2$) has been obtained 
by Alastuey and Jancovici\cite{AlaJanco}:
\beq
   F_{\rm diff} = N_e k_B T \,{\Gamma^2\over 8r_s}
    \left[{2\over u\,\tanh u} - {2\over u^2} + {1\over3}\right],
\label{WK}
\eeq
with $u$ defined as in \req{mu_e}.
The expression in the square brackets in \req{WK} goes to 1 at $u\to0$,
recovering the well-known zero-field result, 
and to $\case13$ at $u\gg1$, reflecting
the fact that two of three degrees of freedom
for the electron motion are frozen out in the strongly
quantizing field. Equation (\ref{WK}) is valid in the
low-density regime, where $r_s\gg\max(\Gamma,\Gamma^{-1})$.
In this regime the correction (\ref{WK}) is smaller than the
classical OCP corrections to the Debye-H\"uckel formula. 
In the electron-ion plasmas, $F_{\rm diff}$ is 
canceled 
because of the local neutrality relation\cite{Cornu}.

In this paper, we are mainly interested 
in the case where 
$(\Gamma r_s)^{-1}\approx 3.16\,T_6 \gtrsim1$.
In this case, a high-temperature expansion \cite{high-T},
which can be written as an expansion in powers of two small parameters
$s_1=\sqrt{\Gamma}/r_s$ and $s_2=\sqrt{\Gamma r_s}$,
is relevant at low densities. The lowest-order correction is the
Hartree-Fock term $\propto\hbar^2 e^2$.
Steinberg\etal\cite{Steinberg} have recently obtained
an analytic result for this term in a magnetic field:
\beq
   { \beta F_{\rm HF}\over N_e } = - {3\Gamma^2\over8r_s} f_1(u),
\label{HF}
\eeq
where the function
\beq
   f_1(u) = {\cosh(2u)\over \cosh^2u}
          \left[ {\tanh u \over u } \right]
          {{\rm arctanh}\left(\sqrt{1-u^{-1}\tanh u}\right)
        \over \sqrt{1-u^{-1}\tanh u}}
\eeq
goes to 1 at small $u$, reproducing the zero-field result\cite{high-T},
and to $\ln (4u)/u$ at very large $u$.

Steinberg\etal\cite{Steinberg} 
have also calculated the corrections
$\propto \hbar e^4$ (the Montroll-Ward and exchange terms).
For the electron gas, they can be written in the form
\beq
   {\beta F_4\over N_e} = {3\sqrt\pi\over16}\,
         {\Gamma^{5/2}\over\sqrt{r_s}}\,
          [ f_2^{ee}(u)+f_3^{ee}(u)\ln2 ],
\label{e4}
\eeq
where $f_2^{ee}(u)$ and $f_3^{ee}(u)$ go to 1 at $u\to 0$,
reproducing the zero-field result\cite{high-T},
and decrease at large $u$.

In order to incorporate these results
into the analytic free-energy model, we employ a simple scaling
procedure. In the fit for $f_{ee}(\theta,\Gamma)=\beta F_{ee}/N_e$
derived in\cite{IIT}, 
where $\theta=T/T_F$ is the degeneracy parameter,
we replace the zero-field value $\theta_0=2\,(9\pi/4)^{-2/3} r_s/\Gamma$
by
\beq
   \theta^\ast = \theta_0\, { 1+\theta_m/\theta_0 
             \over  1+f_1(u)\,(\theta_m/\theta_0)\,\exp(-\theta_m^{-1}) }.
\label{scaling}
\eeq
Here, 
$\theta_m=8\,\gamma^2 r_s^5/(9\pi^2\Gamma)=0.166\,\theta_0\,\gamma^2 r_s^4$
is the value of the degeneracy parameter in the strongly quantizing field.

The scaling (\ref{scaling}) reproduces limiting cases:
({i}) at $r_s\gg1$, the classical OCP expression is recovered,
independently of other parameters;
({ii}) in the nonquantizing regime $\gamma r_s^2\ll1$,
we get the nonmagnetic value $\theta^\ast=\theta_0$;
({iii}) in the strongly quantizing degenerate regime
$\gamma r_s^2\gg1$ and $\theta_m\ll1$,
the correct value of the degeneracy parameter $\theta^\ast=\theta_m$
is recovered;
and
({iv}) in the strongly quantizing nondegenerate regime
$\gamma r_s^2\gg1$ and $\theta_m\gg1$,
the fit reproduces \req{HF} in its range of validity.

\begin{figure}
\epsfysize=70mm
\epsffile[60 200 520 630]{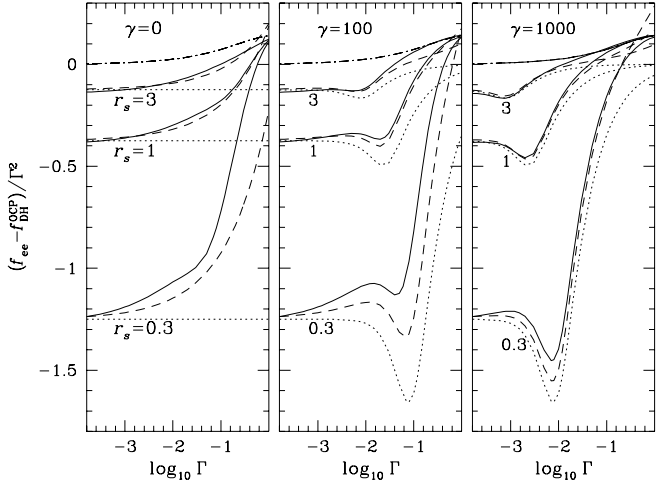} 
\caption{
Contribution to the electron-gas nonideal free energy
$f_{ee}=\beta F_{ee}/N_e$ beyond 
$f_{\rm DH}^{\rm OCP}=-\Gamma^{3/2}/\sqrt3$
 at three values of the magnetic field parameter $\gamma$
 and three values of the density parameter $r_s$ (indicated).
 The scaled fit (solid lines; see text) is compared with the
 high-temperature expansion up to orders $e^2$ and $e^4$
 (dotted and dashed lines, respectively)
 and with the classical OCP (dot-dashed lines).
 }
 \label{fig-ee}
\end{figure}

Figure~\ref{fig-ee} demonstrates the validity of the adopted
modification of $F_{ee}$ at $\Gamma<1$, for three values of $r_s$,
for which the quantum contributions to $F_{\rm ex}^{C}$ could
appreciably affect our results.
We plot {\em departures\/} of $F_{ee}$ from the OCP Debye-H\"uckel
function $F_{\rm DH}^{\rm OCP}$, 
normalized to $N_e k_B T$
and divided for convenience by $\Gamma^2$.
The dot-dashed line shows the classical OCP free energy\cite{CP98},
the dotted line displays the $e^2$ correction (\ref{HF}),
and the dashed line results from inclusion 
of the $e^4$ corrections (\ref{e4}).

The left panel presents the nonmagnetic case.
The solid line shows the fit to $F_{ee}$\cite{IIT}.
The region of approximate coincidence of the fit with 
the high-$T$ expansion can be adopted as the region 
of validity of the latter.
At large $r_s$, it is restricted by the condition
$s_2\ll1$ ($\Gamma\ll r_s^{-1}$), 
while at small $r_s$, the condition $s_1\ll1$
($\Gamma\ll r_s^2$) is more restrictive.

The middle and right panels show the modifications of $F_{ee}$
at two values of the magnetic field strength.
One can see that the scaled fit satisfactorily reproduces
the expansion in the validity range of the latter.
Surprisingly, although the scaling is based on the lowest-order
$e^2$ term (\ref{HF}), the $e^4$ terms
(\ref{e4}) are also well reproduced.

For the electron-ion plasma, the screening contribution $F_{ie}$
should be taken into account. 
At $B=0$, it has been calculated in a number of papers
(e.g., Refs.\cite{IIT,C90,CP98}) 
and fitted by analytic expressions\cite{CP98}.
In a strongly quantizing magnetic field,
this contribution 
has been analytically
evaluated only for a dense plasma at zero temperature using
the linear response theory\cite{Fushiki}.
Comparison with 
the analogous zero-field result\cite{Salpeter}
shows that the strongly quantizing magnetic field increases
the screening energy at high density by factor
$0.8846\,\gamma^2 r_s^4$. 
To our knowledge, there were no 
relevant calculations at arbitrary degeneracy.
In the regime of low degeneracy and weak Coulomb coupling, 
integral representations of the low-density expansion 
coefficients have been obtained\cite{Steinberg,Cornu}.
The contribution of order $e^4$ is given by
Eqs.~(20), (21) of Ref.\cite{Steinberg}.
It is reproduced if to replace $r_s$ by $r_s/(f_2^{ep})^2$
in the nonmagnetic expression\cite{high-T}.
Here, $f_2^{ep}$ can be approximated (within 0.5\%) as
\beq
    f_2^{ep} = \frac12 + 
        t^{0.9} \, {{\rm arctanh}\left[(1-t)^{0.6}\right]
                   \over 2\,(1-t)^{0.6}},
\label{f2}
\eeq
where $t\equiv\tanh(0.4u)/(0.4 u)$.
We apply the scaling $r_s\to r_s/(f_2^{ep})^2$
to the formula for $F_{ie}(r_s,\Gamma)$ given in \cite{CP98}.

\section{Partially ionized plasma}
\label{sect-PI}
\subsection{Hydrogen bound species in the strong magnetic field}
\subsubsection{Atoms}
\label{sect-atoms}
Only a brief summary of the properties of the hydrogen atom
in a strong magnetic field is given below. See, e.g., Ref.\cite{P94}
for details and references. 

The motion of an atom in a magnetic field ${\bf B}$
can be conveniently described using the pseudomomentum 
${\bf K}$, the quantum-mechanical constant of motion related to the 
average center-of-mass velocity  
${\bf v}=\nabla_{\bf K}{E}$, 
where ${E}$ is the total energy of the atom.
If there were no Coulomb attraction,
the energy would be
$
     {E}=E^\perp_{\NL s}+K_z^2/(2\mH),
$
where
\beq
   E^\perp_{\NL s} = 
   \NL\hbar\omega_c+(\NL+s)\hbar\omega_{cp}
\label{Eperp}
\eeq
is the energy of the transverse excitation,
$\NL$ is the electron Landau number,
$s$ is the $z$ projection of the relative proton-electron 
angular momentum,
and $K_z^2/(2\mH)$ is the kinetic energy of motion along the field.
The Coulomb interaction mixes the Landau orbitals.
Nevertheless, it is convenient to keep 
the quantum numbers $\NL$ and $s$ 
for enumerating the quantum states at $\gamma\gg1$.
Then the energy of the atom can be decomposed as follows:
\beq
    {E}_{\NL s\nu}({\bf K}) = E^\|_{\NL s\nu}(\kp) + 
    E^\perp_{\NL s} + K_z^2/(2\mH), \label{2.1} 
\eeq
where
$E^\|_{\NL s\nu}(\kp)<0$ is the ``longitudinal'' energy,
and the quantum number $\nu$ enumerates the longitudinal excitations. 
At $\gamma\gg1$, the states with 
$\NL\neq0$ and large $s$ are subject to autoionization.
Therefore we put $\NL=0$
and suppress this quantum number hereafter.
The binding energy is 
\beq
   \chi_{s\nu}(\kp) = | E^\|_{s\nu}(\kp) | - s\hbar\omega_{cp}. 
\label{chi}
\eeq
 
In accordance with Sec.~\ref{sect-protons},
the zero-point and spin terms are subtracted from \req{Eperp}
and absorbed into \req{DeltaF}.
Note that \req{DeltaF} is valid for the partially ionized 
plasma provided that the IPFs for atoms with opposite proton 
spin projections are identical.
It is true under the assumption that 
the autoionization processes with
proton spin flip may be neglected on the plasma
relaxation time scale.
We adopt this assumption, because the plasma under consideration
is rather dense and nonrelativistic.
Otherwise, states with binding energy
$\chi_{s\nu}(\kp)< g_p\hbar\omega_{cp}/2$ 
should be excluded from the IPF for atoms with the negative 
proton spin projection.

At $K=0$, the atom is axially symmetric, and
its sizes transverse to the magnetic field
can be approximated\cite{CanutoVentura,P98}
as $l_x=l_y\approx a_m\,\sqrt{s+1}$, 
while 
the longitudinal size is much larger: $l_z\sim a_0/\ln\gamma$
for the {\em tightly bound\/} states ($\nu=0$)
and $l_z\sim a_0\nu^2$ for the {\em hydrogenlike\/} states ($\nu\geq1$).
Longitudinal energies of the former states grow
as $E^\|\propto(\ln\gamma)^2$,
whereas the energies of the latter states are relatively small,
$|E^\| | \sim (e^2/a_0)(2n^2)^{-1}$,
where $n$ is the integer part of $(\nu+1)/2$.

An atom moving across the field 
acquires a constant dipole moment
in the direction
opposite to its {\it guiding center\/}
$
{\bf r}_c = c (eB^2)^{-1} {\bf B}\times{\bf K}.
$
When $\kp$ is small enough, the dipole moment is also small, and 
$E^\|$ is increased by $\kp^2/(2m^\perp_{s\nu})$. Here,
 $m^\perp_{s\nu}$ is the 
so-called {\it effective transverse mass}, 
which exceeds $\mH$ and grows with 
field strength. In this case the 
average transverse velocity is $v_\perp=\kp/m^\perp_{s\nu}$.
When $\kp$ exceeds some critical value 
$K_c\sim10^2\hbar/a_0$, the atom becomes {\it decentered}: 
$v_\perp$ reaches a maximum and starts to decrease, while the 
electron-proton separation approaches $r_c$.
Thus, for the decentered states, the transverse pseudomomentum $\kp$
characterizes electron-proton separation
rather than velocity.

In the limiting case of 
$\kp\gg\gamma(\nu+\frac12)^2\hbar/a_0$, 
the longitudinal energies approach 
the asymptote $E^\|\sim -e^2/r_c$. 
Note that only the states with $s=0$ may remain bound
if they have such large values of $\kp$.
Indeed, since $E^\|$ is small for large $\kp$,
the binding energy (\ref{chi}) becomes negative for $s\geq1$.
However, at $s=0$ and arbitrarily large $\kp$,
there still remains an infinite number of truly bound states
(enumerated by $\nu$), as has been strictly proved in \cite{P94}.

Since $r_c=a_0^2\kp/\gamma\hbar$, the decentered states
have huge sizes at $\gamma < 1$; hence they are expected to be destroyed
by collisions with surrounding particles
in the laboratory and in white-dwarf atmospheres
\cite{positronium}.
 In neutron-star atmospheres
at $\gamma\gtrsim10^3$, however, the decentered states
may be significantly populated, as we shall see below.

Accurate numerical dependences of the atomic binding energies
$\chi_{s\nu}(\kp)$ and sizes $l_{s\nu}(\kp)$
for $300\leq\gamma\leq10^4$ and any $\kp$ 
have been obtained in \cite{P94,P98}.
In Ref.\cite{P98}, analytic fits have been constructed
for these quantities as functions of $\gamma$ and $\kp$,
as well as for the critical pseudomomentum $K_c$
and transverse mass $m^\perp_{s\nu}$ as functions of $\gamma$,
for various $s$ and $\nu$. These fits are more suitable for
studying the thermodynamics of hot plasmas
than previous approximations\cite{LS95}, which were accurate
only for the centered ground state ($s=\nu=0$ and $\kp<K_c$).
\subsubsection{Other bound species}
\label{sect-molecules}
At sufficiently low $T$ or high $B$, there may exist a considerable
amount of molecules and ions in the hydrogen plasma.
The molecular ion ${{\rm H}_2}^+$ 
has been thoroughly investigated at $B<10^{10}$~G,
including the dependence of binding energies
of various electron-vibrational-rotational levels
on the angle between the ion axis and 
the magnetic field direction\cite{Wille}.
For stronger fields, only the parallel configuration 
has been considered\cite{Kher,Neu,LS96,Kravchenko}.
The ions ${{\rm H}_2}^+$ have negligible abundance in the strong
field, owing to the small binding energy, compared to 
the atoms and H$_2$ molecules\cite{Kher,LS97}.
The same is probably true for the H$^-$ ions\cite{LS97}.

H$_2$ molecules have been studied in detail at various 
field strengths\cite{Neu,LS96,Detmer}. 
An interesting result is that the
ground state is unbound at $0.18<\gamma<12.3$\cite{Detmer}.
Fitting formulas for the dissociation energies
in the parallel configuration for $\gamma\gtrsim10^3$
have been given in\cite{LS97,LS96}.
At such fields, the dissociation energy grows
$\propto(\ln\gamma)^2$,
approximately at the same rate as the atomic 
ground-state energy.
The equilibrium internuclear distance decreases as
$1/\ln\gamma$, being as small as $\case14 a_0$ at $B=10^{12}$~G,
again roughly proportional to the longitudinal size of the atom.

Moreover, strong magnetic fields stabilize polymer chains
H$_N$ aligned with ${\bf B}$, 
as first suggested by Ruderman\cite{Ruderman} 
and later confirmed by Hartree-Fock calculations\cite{Neu}.
The specific quantum-mechanical properties of these species
(e.g.\ their excitation spectra) are poorly known.

Motional effects on the molecules and chains in the strong magnetic fields
have not been studied. 
Therefore, one cannot construct a reliable EOS
in the domain of $\rho, T, B$ 
where these species are expected to dominate.
For instance, Lai and Salpeter\cite{LS97} estimated 
the effective transverse mass of H$_N$ as $N$ times
the atomic effective mass, $N m^\perp_{00}$,
and used it in the dissociation equilibrium equation.
However, since the heavier molecule has lower velocity
at a given $K$, it is exposed to a weaker electric field
in the comoving frame. Therefore, one could expect
its energy levels to be less perturbed and its
effective mass to be closer to the zero-field
value, $N m_{\rm H}$.

Because of these uncertainties, we do not include H$_N$
in our study but restrict ourselves to the atomic phase.
Nevertheless, we include ground-state H$_2$ molecules
in order to determine the
validity domain of our results.

\subsection{Free-energy model}
Our free-energy model is a straightforward generalization
to the magnetic case of the model presented 
in Sec.~\ref{sect-zero-model}:
\beq
   F = F_{\rm id}^{(e)} + F_{\rm id}^{(p)} + F_{\rm id}^{\rm neu}
      +F_{\rm rad}
       + F_{\rm ex}^{C} + F_{\rm ex}^{\rm neu}.
\label{Fren}
\eeq
The ideal electron and proton free energies $F_{\rm id}^{(e)}$
and $F_{\rm id}^{(p)}$ are derived in Secs.~\ref{sect-electrons}
and \ref{sect-protons}, respectively.
$F_{\rm rad}$ is given by \req{Frad}.
The Coulomb part $F_{\rm ex}^{C}$ has been discussed
in Sec.~\ref{sect-Coul}. Now let us consider
the ideal and nonideal contributions $F_{\rm id}^{\rm neu}$
and $F_{\rm ex}^{\rm neu}$ brought about by the bound species.

Since the quantum-mechanical characteristics of an atom
 in a strong magnetic field 
depend in a nontrivial way on the transverse pseudomomentum $\kp$,
the distribution of atoms over $\kp$ cannot be written in a closed form,
and only the distribution over
$K_z$ remains Maxwellian.
Let $p_{s\nu}(\kp)\dd^2\kp$ be the probability to find
an atom with given $(s,\nu)$
in an element $\dd^2\kp$ near the point $\kpvec$
of the transverse pseudomomentum plane. 
For the Maxwell distribution, we would have 
$p_{s\nu}(\kp)=(2\pi\hbar)^{-2}\lambda_{\rm H}^2\exp[-\kp^2/(2m_{\rm H})]$.
In the general case, the number
of atoms in an element $\dd^3K$ of the pseudomomentum space
is
\beq
  \dd N({\bf K})  = N_{s\nu}\,
        {\lambda_{\rm H} \over2\pi\hbar}\,
         \exp\left(-{\beta K_z^2\over 2\mH}\right)\, p_{s\nu}(\kp) \dd^3K,
\eeq
where $N_{s\nu}=\int\dd N_{s\nu}({\bf K})$ 
is the total number of atoms with the
specified discrete quantum numbers.
The distribution $N_{s\nu} p_{s\nu}(\kp)$ is not given in advance
but should be calculated self-consistently by minimization of 
the total free energy, including nonideal terms.

It is convenient to introduce deviations
from the Maxwell-Boltzmann distribution through the occupation 
probabilities $w_{s\nu}(\kp)$:
\bea
  p_{s\nu}(\kp) & =  & 
       \left( { \lambda_{\rm H}\over 2\pi\hbar } \right)^2
        { w_{s\nu}(\kp)\exp[\beta\chi_{s\nu}(\kp)]
      \over Z_{s\nu} },
\label{psn}
\\
   N_{s\nu}/N_{\rm H}  & = &  Z_{s\nu}/Z_w,
\label{Nsn}
\eea
where
\bea
   Z_{s\nu}  & = &  
{ \lambda_{\rm H}^2\over 2\pi\hbar^2 } 
          \int_0^\infty w_{s\nu}(\kp){\rm e}^{\beta\chi_{s\nu}(\kp)}
            \kp \dd \kp ,
\label{Zsn}
\\
  Z_w  & = &  \sum_{s\nu} Z_{s\nu}.
\label{Zw}
\eea

The number of atoms per unit phase-space cell
equals $[\dd N({\bf K})/\dd^3K](2\pi\hbar)^3/V$.
Calculation of $(U-TS)$
for this distribution gives
\bea
&&
   F_{\rm id}^{\rm H} = k_B T \sum_{s\nu} N_{s\nu}
    \int \left\{
\ln\left[n_{s\nu} \lambda_{\rm H} (2\pi\hbar)^2 p_{s\nu}(\kp)\right]
\right.
\nonumber\\&&\qquad\left.
       \vphantom{)^2}
        - 1 - \beta\chi_{s\nu}(\kp) \right\} p_{s\nu}(\kp)\dd^2\kp
\label{FH1}
\\&& = k_B T \sum_{s\nu} N_{s\nu}
     \int\ln\left[n_{s\nu} \lambda_{\rm H}^3 
        {w_{s\nu}(\kp) \over \exp(1) Z_{s\nu}}\right]
       p_{s\nu}(\kp)\dd^2\kp.
\nonumber
\eea
The contribution of molecules should be added to this expression.
We estimate it taking into account only the molecules
in their ground state. 
This is an acceptable approximation at $B\gtrsim10^{12}$~G,
because the energies of different types of molecular excitations
are not much smaller than the electronic excitations
of the atoms\cite{LS96} (contrary to the zero-field case), 
so that excited levels cannot give a large
contribution to the molecular IPF at those relatively low
temperatures where the molecular fraction is large\cite{LS97}.
We also neglect the (unknown) motional modification of the molecular
spectrum, as noted in Sec.~\ref{sect-molecules}.

Thus, we include in the ideal free energy of the bound species 
$F_{\rm id}^{\rm neu}$ the term 
\beq
   F_{\rm id}^{{\rm H}_2} = 
       N_{{\rm H}_2} k_B T \,[ \ln(n_{{\rm H}_2}\lambda_{{\rm H}_2}^3)
        -1 -\chi_{{\rm H}_2} ],
\eeq
where $\chi_{{\rm H}_2}=2\chi_{00}(0)+Q_2$ is the molecular binding energy,
and $Q_2$ is the dissociation energy
fitted as function of $\gamma$ in\cite{LS97,LS96}.

The nonideal part $F_{\rm ex}^{\rm neu}$ is calculated in the
hard-sphere approximation using Eqs.~(\ref{F_HS})--(\ref{diam}),
where the composite atomic number is $\kappa=(s\nu\kp)$
and the obvious generalization of
$\sum_\kappa$ includes $\int p_{s\nu}(\kp)\dd^2\kp$. 
The effective atomic size $l_\kappa = l_{s\nu}(\kp)$
is given by fitting formulas\cite{P98}.
The effective size of the H$_2$ molecule
in the ground state is estimated as 
 $l_{{\rm H}_2} = [2a_m^2 + l_{z,{\rm H}_2}^2]^{1/2}$,
where the longitudinal size 
is $l_{z,{\rm H}_2}\approx l_{z0}+r_0$.
Here, $l_{z0}$ 
is the longitudinal size of the ground-state H atom 
fitted in \cite{P98},
and $r_0\approx12.7\,(\ln\gamma)^{-2.2}$ is the equilibrium
internuclear separation given in \cite{LS97}.
\subsection{Equilibrium conditions}
The thermodynamic equilibrium for the free-energy model (\ref{Fren})
is given by a generalization of 
the equations in Sec.~\ref{sect-Saha0}, taking into account
the fact that the atomic IPF $Z_w$ now includes integration over $\kp$.
In particular, $\partial F_{\rm ex}/\partial N_\kappa$
in \req{occprob0} is replaced by a functional derivative.

In the conditions studied here,
neutral atoms can exist only in the regime of 
strong magnetic quantization and weak degeneracy. 
Therefore it is convenient to write the generalized Saha
equation using \req{Fe-magn}
and describe the deviations
from it by a separate factor $\Lambda$. 
For the ideal free energy
of protons, we use \req{Fp}, and the one for the atoms 
is given by \req{FH1}. Thus the generalized Saha equation
reads
\beq
    n_{\rm H} =  n_p n_e                                           
   { \lambda_p\lambda_e (2\pi a_m^2)^2
     \over \lambda_{\rm H}^3 } 
    \, \left[1-\exp(-\beta\hbar\omega_{cp})\right] \, 
    Z_w {\rm e}^{\Lambda},
\label{Saha}
\eeq
where 
\beq
   \Lambda=
        \beta \mu_e  - \ln(2\pi a_m^2\lambda_e n_e)
    + \beta\left( {\partial \mu_e\over\partial\ln n_e} - 
        {\partial P_e \over \partial n_e} \right)  
\eeq
allows for deviations of the exact value of 
$F_{\rm id}^{(e)}$ from that given by \req{Fe-magn}
due to electron degeneracy and population of excited Landau levels.
The distributions of atoms over the discrete quantum numbers
and over the transverse pseudomomenta are given by Eqs.~(\ref{Nsn})
and (\ref{psn}), respectively.

The occupation probabilities can be presented as a product of 
two terms that arise 
from $F_{\rm ex}^C$ and $F_{\rm HS}$:
\beq
   w_{s\nu}(\kp)= w^C w_{s\nu}^{\rm HS}(\kp).
\label{occprob}
\eeq
Hereafter, we exclude $N_e$ from our formulas
by explicit use of the electroneutrality condition $N_e = N_p$.
Then the Coulomb factor reads
\beq
   \ln w^C = \beta {\partial  F_{\rm ex}^C \over \partial N_p}
      = 2 f_{\rm ex}^C 
      + \frac23 \left( {\partial f_{\rm ex}^C \over \partial \ln\Gamma}
      - {\partial f_{\rm ex}^C \over \partial \ln r_s} \right),
\eeq
where $f_{\rm ex}^C(r_s,\Gamma)\equiv \beta F_{\rm ex}^C / (2N_p)$
is described in Sec.~\ref{sect-Coul}.
In the Debye-H\"uckel limit, 
$w^C$ is given by \cite{P96}
\beq
       \ln w^C_{\rm DH} = - \sqrt{8\pi n_p (\beta e^2)^3}.
\eeq
The hard-sphere factor reads
\beq
      \ln w^{\rm HS}_{s\nu}(\kp) = {
    (1-\eta/2)\ln w^{(0)}_{s\nu}(\kp) -5\eta^2+3\eta^3 
    \over (1-\eta)^3},
\eeq
where $\ln w^{(0)}=(\partial/\partial N_p 
- \partial/\partial N_\kappa)[4N_{\rm tot}\eta]$ 
is the low-density limit of $\ln w^{\rm HS}$
and $\eta$ is the packing fraction (\ref{eta}).
Explicitly,
\bea
    \ln w^{(0)}_{s\nu}(\kp) & = & - {4\pi\over3} \left\{
       (n_{\rm H} + n_p) l_{s\nu}^3(\kp) 
\right.
\nonumber\\
    & & + n_{{\rm H}_2}[l_{s\nu}(\kp)+l_{{\rm H}_2}]^3
\nonumber\\
    & &   + \left. 3 n_{\rm H} [ l_{s\nu}(\kp) \langle l^2 \rangle
            + l_{s\nu}^2(\kp) \langle l \rangle ] \right\},
\eea
\bea
   \eta & = & {\pi\over 3N_{\rm tot}V } \left[
       N_{\rm H}^2 (\langle l^3 \rangle 
            + 3 \langle l^2 \rangle \langle l \rangle)
         + N_{\rm H} N_p \langle l^3 \rangle
\right.
\nonumber\\
     & &    + N_{\rm H} N_{{\rm H}_2} ( \langle l^3 \rangle
         + 3 \langle l^2 \rangle\, l_{{\rm H}_2}
         + 3 \langle l \rangle\, l_{{\rm H}_2}^2 + l_{{\rm H}_2}^3 )
\nonumber\\
     & & + \left. N_p N_{{\rm H}_2} l_{{\rm H}_2}^3 +4 N_{{\rm H}_2}^2 l_{{\rm H}_2}^3 \right],
\eea
where
\bea
   \langle l^k \rangle & \equiv & {1\over N_{\rm H}}\sum_{s\nu} N_{s\nu}
         \int l_{s\nu}^k(\kp) p(\kp) \dd^2 \kp.
\eea

The dissociation equilibrium is given by \req{disseq},
where 
 $Z_{w2}$ is replaced by $w_{{\rm H}_2}\exp(\chi_{{\rm H}_2})$
and
$Z_w$ is modified according to Eqs.~(\ref{Zsn}) 
and (\ref{Zw}). From \req{wH2} we obtain
$w_{{\rm H}_2} = (w^C)^2 w_{{\rm H}_2}^{\rm HS}$, where
\bea
   \ln w_{{\rm H}_2}^{\rm HS} & = & {
    (1-\eta/2) \ln w^{(0)}_{{\rm H}_2} -15\eta^2+9\eta^3 
    \over (1-\eta)^3},
\\
    \ln w^{(0)}_{{\rm H}_2} & = & - {4\pi\over3} \left[
       n_{\rm H} (3 \langle l^2 \rangle\, l_{{\rm H}_2}
         + 3 \langle l \rangle\, l_{{\rm H}_2}^2 - \langle l^3 \rangle)
\right.
\nonumber\\
       && +  \left. (n_{\rm H} + n_p + 6 n_{{\rm H}_2}) l_{{\rm H}_2}^3 \right].
\label{w0H2}
\eea

\begin{figure}
 \epsfysize=63mm
 \epsffile[70 300 400 700]{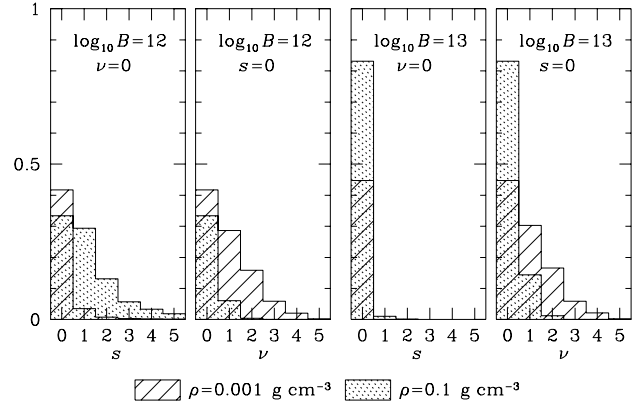} 
 \caption{Distribution of atomic occupation numbers
 at $T=10^6$~K for the magnetic field strengths $B=10^{12}$~G
 and $10^{13}$~G (indicated).
 For each value of $B$, the distribution over 
 the quantum number $s$ at $\nu=0$ and over $\nu$
 at $s=0$ is shown for two density values:
 $\rho=0.001\gcc$ (hatched histograms) and $0.1\gcc$
 (shaded histograms).
 }
 \label{fig-distr}
\end{figure}

A solution of Eqs.~(\ref{Saha})--(\ref{w0H2}),
supplemented by the stoichiometric constraint
$n_p + n_{\rm H} + 2n_{{\rm H}_2} = n_0$, yields
the equilibrium abundances and the free energy value.
The solution is sought by an iteration procedure,
in analogy with the zero-field case
described in Sec.~\ref{sect-Saha0},
and the EOS is obtained from \req{TDrel}.
In the strongly quantizing magnetic field and in the 
nondegenerate regime, the EOS
is a sum of three analytic terms: the ideal term
$P_{\rm id}=n_{\rm tot} k_B T + (4\sigma/3c)\, T^4$,
the contribution due to the Coulomb nonideality
given by derivation of the fit described in Sec.~\ref{sect-Coul},
and the hard-sphere contribution
$P_{\rm HS}= 4\eta\,(1-\eta/2)(1-\eta)^{-3} n_{\rm tot} k_B T$.
\section{Results}
\label{sect-res}
\subsection{Distribution of plasma particles}
\label{sect-distrib}
\subsubsection{Occupation numbers}
Figure~\ref{fig-distr} displays the distribution of the atoms over 
quantum states given by \req{Nsn} at $B=10^{12}$~G and $10^{13}$~G,
$T=10^6$~K, and
at two relatively low densities $\rho=0.001\gcc$
and $0.1\gcc$. 
The left panel shows the relative occupation numbers
for the tightly bound states $\nu=0$,
for different quantum numbers $s$. 
The distribution is broader for higher density.
This apparently surprising feature
is easily explained by the presence of 
the third quantum parameter $\kp$,
in addition to $s$ and $\nu$.
At low density most atoms reside in the states with 
large values of $\kp$
because of the large statistical weight of such states,
which all have $s=0$ (Sec.~\ref{sect-atoms}).
At higher density, these strongly decentered states 
are removed by the excluded-volume effects,
and the distribution over $s$ grows broader.
Conversely, on the neighboring panel
we observe a narrower distribution over $\nu$
at higher density, because the excluded-volume
effects eliminate the hydrogenlike states.
In the next section we shall see that ultimately,
at still larger densities,
only the ground centered state survives
($s=\nu=0$, $\kp<K_c$).

The right two panels demonstrate the effect of 
increasing $B$ to $10^{13}$~G.
Due to the larger binding energies,
the distribution at $\rho=0.1\gcc$
has become narrow, with more atoms concentrated
in the ground state. However, at the lower density,
the distribution over $\nu$ has changed weakly,
since the increase of binding energies
is accompanied by a decrease of the atomic size
(and hence a decrease of the nonideality effects).

\subsubsection{Ionization equilibrium}

\begin{figure}
 \epsfysize=120mm
 \epsffile[105 165 690 710]{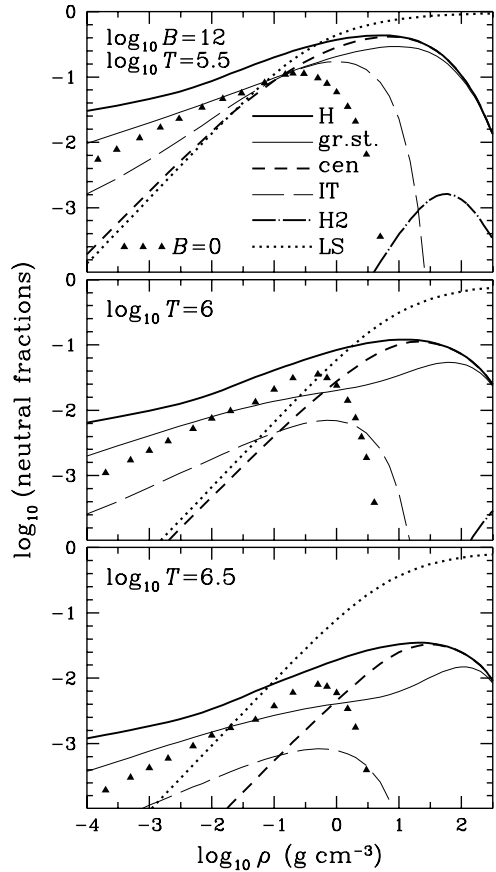} 
 \caption{
 Ionization isotherms at $B=10^{12}$~G and three values of $T$
 (indicated):
 total fraction of atoms $f_{\rm H}=n_{\rm H}/n_0$
 (heavy solid lines) and the fractions of ground-state atoms
 (thin solid lines), the centered atoms (short-dashed lines),
 and the optically identifiable (Inglis-Teller) atoms
 (long-dashed lines).
 The dot-dashed lines 
 show the molecular fraction $f_{{\rm H}_2}=2n_{{\rm H}_2}/n_0$,
 which is below the frame in the bottom panels.
 For comparison, $f_{\rm H}$ in the 
 zero-field case (triangles)
 and in the approximation of Lai and Salpeter
 (dotted lines) is also shown.
 }
 \label{fig-ie1}
\end{figure}

Figure~\ref{fig-ie1} shows the ionization curves at 
three values of $T$ for $B=10^{12}$~G.
The heavy solid lines represent the total fraction
of atoms $f_{\rm H}=n_{\rm H}/n_0$ in all quantum states,
calculated according to \req{Saha}.
Thin solid lines show the fraction $f_{00}$ of atoms in the ground
state ($s=\nu=0$, but any $\kp$), and
the dashed lines show the fraction 
of atoms in the centered states
($\kp<K_c$, any $s$ and $\nu$).
For reference, triangles display the zero-field
atomic fraction given by \req{Saha0}.

We see that a strong magnetic field generally increases
the neutral fraction. 
At low densities, the excited atoms
contribute significantly. 
Since their effective size is proportional to $\kp$,
the integration (\ref{Zsn}) gives roughly 
$Z_{s\nu}\propto n_0^{-2/3}$;
therefore $f_{00}$ decreases asymptotically as
$n_0^{1/3}$. Because of the broadening
of the $\nu$ distribution
(roughly, $\max\nu\propto n_0^{-1/6}$), 
the low-density wing of the curve for the total neutral fraction
has a slope $f_{\rm H}\propto n_0^{1/6}$,
which is very moderate compared to $f_{\rm H}\propto n_0^{1/2}$
in the nonmagnetic case (triangles).

The centered atoms,
whose pseudomomentum is limited from above by 
the critical value $K_c$, 
have a nearly density-independent IPF at low $\rho$.
Therefore their fraction behaves as $f_{\rm cen}\propto n_0$,
and they disappear much faster at low $\rho$
and especially at high $T$ (compare the dashed lines
in the upper and lower panels).

At high densities, on the contrary,
the decentered states become depleted 
due to the excluded-volume effects, so that
the dashed line in the figure merges with the solid one
at $\rho\gtrsim10\gcc$.
At still higher densities $\rho\gtrsim10^2\gcc$,
all excited states disappear, and
only the state $s=\nu=0$ survives.
The pressure ionization
proceeds around $\rho\sim10^2-10^3\gcc$.
The excluded-volume and Coulomb nonideal effects
favor pressure ionization
[both $w^{\rm HS}$ and $w^C$ in \req{occprob} are less than unity],
whereas finite electron degeneracy hampers it ($\Lambda>0$).
Because of the reduced atomic volume,
the pressure ionization
occurs at densities orders of magnitude larger
than for the zero-field case \cite{SC}.
At $T=10^{5.5}$~K, 
the molecular fraction becomes non-negligible at $\rho\sim10^2\gcc$. 

Not all of the neutral atoms that contribute to the EOS
may be identified spectroscopically.
Because of their perturbation by plasma microfields,
the atoms that do not satisfy the Inglis-Teller criterion
form ``optical continuum.''
An approximate estimate of 
the fraction of atoms below the optical continuum
is given by a generalization of the
optical occupation probabilities $\tilde{w}_\kappa$
(Sec.\ \ref{sect-res0})
to the case of the strong magnetic field
according to Eq.~(14) of Ref.~\cite{PP95}. 
This ``IT'' fraction
is shown by the long-dashed lines.
Their rapid decrease indicates 
that the atomic spectral features disappear around
$\rho\sim10\gcc$, long before pressure ionization.

\begin{figure}
 \epsfysize=120mm
 \epsffile[105 165 690 710]{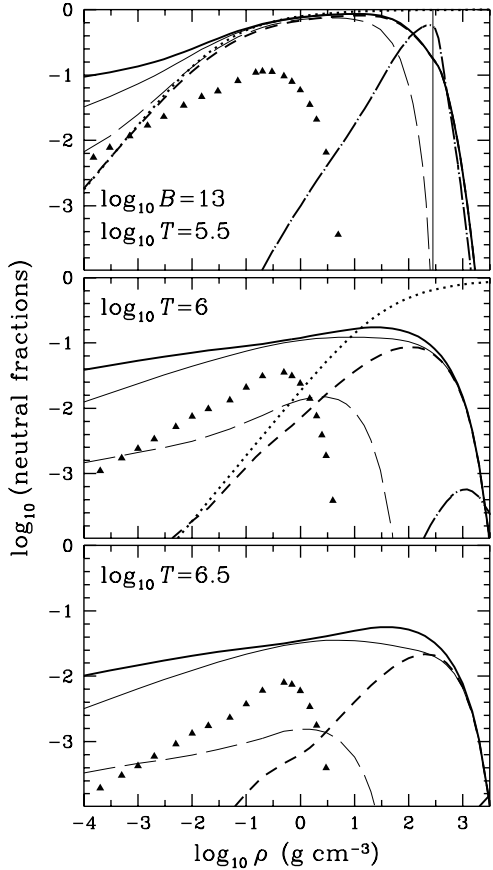} 
 \caption{
 The same as in Fig.~\ref{fig-ie1} for $B=10^{13}$~G.
 The vertical line in the top panel
 separates the region of thermodynamic instability.
 }
 \label{fig-ie10}
\end{figure}

The approximation of Lai and Salpeter\cite{LS97},
also shown in the figure (dotted line), clearly underestimates the neutral
fraction at low density and overestimates it at high
density, especially at high temperature.
At low density, the discrepancy arises mainly
from an underestimation of the decentered 
states because of an incorrect fitting formula
to their binding energies. From comparison 
with the dashed line in Fig.~\ref{fig-ie1}, 
we see that the fraction of centered states can be
estimated by the approximation\cite{LS97}
at $\rho<0.1\gcc$ and $T<10^6$~K.
At higher $T$ or larger $\rho$, the atomic abundance
is overestimated in\cite{LS95,LS97} because of 
neglecting nonideal effects.
Although the neutral fraction is very significant,
it never dominates the plasma
at the values of $T$ and $B$ shown in Fig.~\ref{fig-ie1},
contrary to the prediction of Ref.~\cite{LS97}.
(At $T=10^{5.5}$~K, the maximum is $f_{\rm H}=0.41$ at $\rho\approx5\gcc$.)

Figure~\ref{fig-ie10} shows the ionization curves 
for a stronger field, $B=10^{13}$~G.
Under this condition, the neutral fraction still increases.
At $T=10^{5.5}$~K (top panel), $f_{\rm H}$ exceeds $\case12$
at $\rho > 0.1\gcc$, reaching the maximum of 85\%
at $\rho\approx10$.
Most atoms in this regime reside in the centered ground state.
On the other hand, at $T=10^{5.5}$~K and $\rho\sim10^2-10^3\gcc$,
the molecules are the dominant species; hence
our present model
may be not accurate in this $\rho-T$ domain. 

A comparison with the result by Lai and Salpeter
is not performed for $T=10^{6.5}$~K
(the bottom panel of Fig.~\ref{fig-ie10})
because the approximations (3.11), (3.12) of Ref.\cite{LS97}
yield a negative IPF in this case.

At $T=10^{5.5}$~K and $\rho\gtrsim300\gcc$, 
there appears thermodynamic instability
($\partial P/\partial\rho < 0$) leading to a phase transition.
The stability is recovered at $\rho\gtrsim 8000\gcc$,
where the plasma is fully ionized. 
This phase transition is a complete analogue
to the plasma phase transition (PPT) 
predicted in the zero-field case by several theoretical
models\cite{SC,PPT} but not yet confirmed
in experiment. 
It is caused by a strong Coulomb attraction between
pressure-ionized plasma particles, which contributes
negative pressure that cannot be compensated at low temperature
until the degeneracy sets in.
There is no confidence in the reality of the PPT
because of its model-dependence.
In our case, an additional uncertainty is introduced 
by the simplified treatment of molecules. 

\begin{figure}
 \epsfysize=77mm
 \epsffile[35 160 540 620]{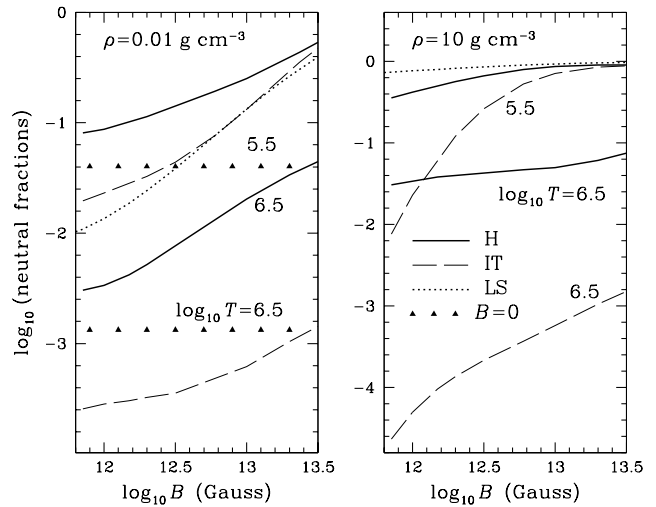} 
 \caption{
 Dependence of the atomic fraction $f_{\rm H}$ (solid lines)
 and the fraction of the optically identifiable atoms
 (dashed lines) on the magnetic field strength
 at two values of $\rho$ (indicated in the figure), 
 $T=10^{5.5}$~K (upper curves) and $10^{6.5}$~K
 (lower ones).
 The atomic fraction at $B=0$ (triangles)
 and the approximation of Ref.\protect\cite{LS97}
 (dotted lines) are shown for comparison.
 }
 \label{fig-ie-B}
\end{figure}

The $B$-dependence of the atomic fraction at two values of $T$
and two values of $\rho$ is shown in Fig.~\ref{fig-ie-B}.
The total $f_{\rm H}$ is drawn by solid lines,
and the ``optical'' (Inglis-Teller) fraction
by dashed lines. Triangles in the left
panel show the total fraction of atoms
at $B=0$ (it is negligible at 
$\rho=10\gcc$ on the right panel).
Dotted lines correspond to the approximation \cite{LS97}
at $T=10^{5.5}$~K.

It was found previously \cite{Gnedin-etal,Kher-1}
that the ionization degree decreases with growing $B$
above $\sim10^{12}$~G only at $T\lesssim5\times10^5$~K
but, in contrast to the present results, 
increases at higher $T$. This behavior was attributed to
two effects: decreasing phase space occupied by a plasma particle
with growing $B$, which favors ionization, and increasing binding energy,
which disfavors it.
Our present result arises from 
the motional perturbations of the atoms,
neglected in \cite{Gnedin-etal,Kher-1}:
first, increasing $B$ increases the effective mass $m^\perp$
and thus the statistical weight of the centered atoms,
and second, at low densities the atomic IPF is further increased 
due to the decentered states.

\subsection{Equation of state}
\label{sect-EOS}
\begin{figure}
 \epsfysize=88mm
 \epsffile[100 200 520 650]{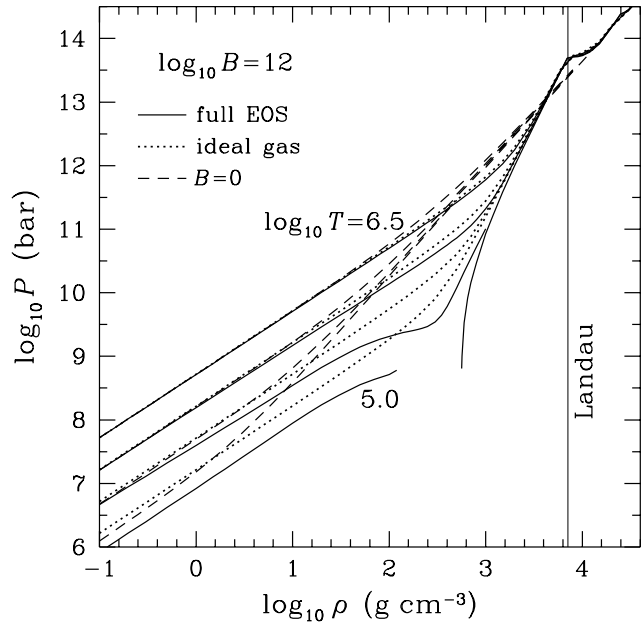} 
 \caption{EOS of partially ionized atomic hydrogen 
 at $B=10^{12}$~G (solid lines)
 compared with the EOS of fully ionized ideal
 electron-proton plasma (dotted lines)
 and the EOS of partially ionized hydrogen at $B=0$
 (dashed lines).
 The temperature logarithms are
 (from top to bottom) $\log T[K] = 6.5, 6.0, 5.5$, and 5.0.
 The vertical line corresponds to $\rho_B$, above which
 excited Landau levels become populated.
 }
 \label{fig-prm1}
\end{figure}

Figure~\ref{fig-prm1} presents four pressure isotherms 
obtained using the free-energy model described in Sec.~\ref{sect-PI}.
For comparison, we also show the fully ionized ideal-gas EOS
(Sec.~\ref{sect-id}) 
and the nonmagnetic EOS (Sec.~\ref{sect-thermo0}).
The vertical line bounds the region $\rho<\rho_B$.
Let us first discuss the low-density regime $\rho\lesssim10\gcc$.
At $T\gtrsim10^6$~K, all three EOS reduce to $P=n_0 k_B T$.
At lower temperatures, the pressure deviates from this 
law because of the partial recombination of atoms.
As discussed in the previous section,
a strong magnetic field increases the neutral fraction;
therefore the pressure is reduced more significantly
compared to the $B=0$ case.

In the intermediate-density range $10\gcc\lesssim\rho\lesssim\rho_B$,
the differences among the three considered cases are most important.
For $B=0$, the plasma is fully ionized in this region,
and the electrons become partially degenerate,
making the EOS stiffer.
In a strong magnetic field,
the electron degeneracy is reduced (Sec.~\ref{sect-FI});
hence the ideal-gas EOS is softer,
except for densities approaching $\rho_B$,
where the degeneracy sets in and pressure
grows rapidly. Partial recombination and Coulomb
nonideality lead to still further decrease of $P$.
The pressure ionization discussed above
has two opposite effects on the pressure:
the positive ideal-gas contribution
of free electrons appearing in the course of the ionization
and the positive nonideal pressure of neutral species 
compete with the negative Coulomb contribution.
At low $T$, these effects may cause the thermodynamic
instability mentioned above, which we observe
on the isotherm $T=10^5$~K.
The second lowest isotherm in the figure is slightly overcritical
for this PPT.
The dependence of the critical temperature $T_c$ 
and density $\rho_c$
on $B$ can be fitted by simple power laws
$T_c=3\times 10^5\,B_{12}^{0.39}$~K
and
$\rho_c = 143\,B_{12}^{1.18}\gcc$,
where $B_{12}\equiv B/(10^{12}$~G).
These fits provide an accuracy of a few percent
in the considered range of the field strengths
$7\times10^{11}{\rm~G} < B < 3\times 10^{13}$~G.

At higher density $\rho\gtrsim\rho_B$,
excited Landau levels become populated due to the 
increase of the Fermi energy.
Eventually, at $\rho\gg\rho_B$, 
the nonmagnetic EOS is recovered.

\begin{figure}
 \epsfysize=65mm
 \epsffile[110 280 330 600]{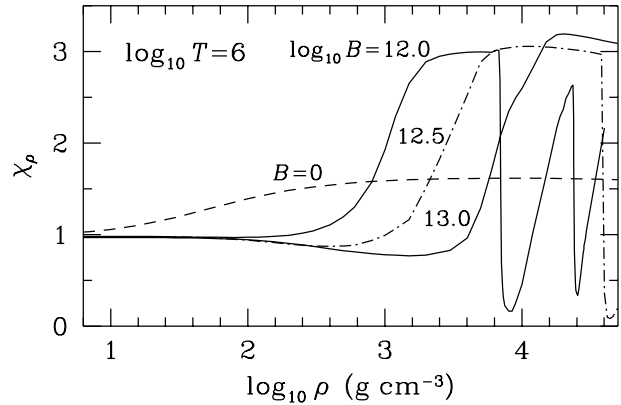} 
 \caption{Density exponent
 $\chi_\rho=(\partial \ln P / \partial \ln \rho)_T$
 at $T=10^6$~K without magnetic field (dashed line)
 and in strong magnetic fields of various
 indicated strengths (dot-dashed and solid lines).
 }
 \label{fig-chir}
\end{figure}

Figure~\ref{fig-chir} demonstrates the effects of
the strong magnetic field on the density exponent
$\chi_\rho=(\partial \ln P / \partial \ln \rho)_T$.
Although
the pressure approaches the nonmagnetic value at $\rho > \rho_B$,
the effects of magnetic quantization remain
quite prominent for the derivative $\chi_\rho$,
as shown by the curve $B=10^{12}$~G in the figure.
Consecutive population of excited Landau levels
causes the oscillations
of $\chi_\rho$ and other second derivatives of $F$
around their nonmagnetic values.
The regime where these oscillations are 
significant is called {\it weakly quantizing}\cite{YaK}.

\begin{figure}
 \epsfysize=65mm
 \epsffile[110 280 330 600]{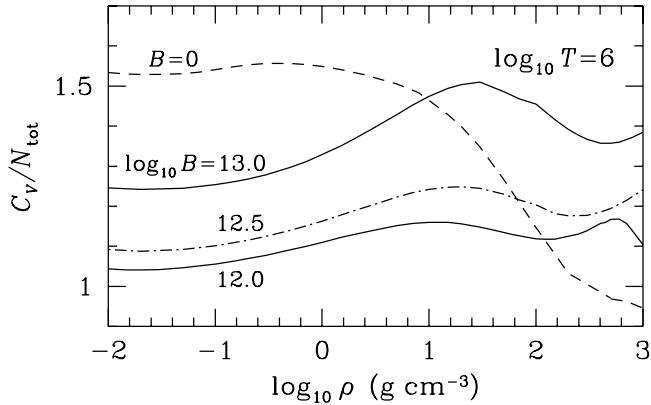} 
 \caption{Normalized heat capacity
 at the same $T$ and $B$ values 
 as in Fig.~\ref{fig-chir}.
 }
 \label{fig-cvm}
\end{figure}

The effects of a strongly quantizing magnetic field on 
the reduced heat capacity 
$C_V=k_B^{-1}(\partial U / \partial T)_V$ 
divided by the number of plasma particles,
$N_{\rm tot}= N_e+N_p+N_{\rm H}+N_{{\rm H}_2}$,
are shown in Fig.~\ref{fig-cvm}.
In the nonmagnetic case (dashed line),
the classical value $C_V/N_{\rm tot}=\case32$
is slightly exceeded at lower densities
because of the thermal ionization of the atoms,
and it is reduced to smaller values at higher densities
because of electron degeneracy.

In strong magnetic fields,  the heat capacity 
is modified due to  several effects.
In the low-density regime, $C_V$
is reduced compared to the nonmagnetic value
because of the quantization of the transverse
motion of electrons and protons.
The strongly quantized electrons have only one 
motional degree of freedom, so that their contribution
reduces to $C_{Ve}=\case12\,N_e$.
When protons are nonquantized and the plasma is fully ionized,
this amounts to $C_V/N_{\rm tot}=1$.
In the general case, the contribution of free spinless protons 
would be
\beq
   { C_{Vp}^{(1)} } = \left[\, {1\over2 }
         + \left( { \beta\hbar\omega_{cp} 
          \over 2\sinh(\case12\beta\hbar\omega_{cp}) } \right)^2 
              \, \right]\,N_p,
\eeq
which tends to $\case12$ at $\beta \hbar\omega_{cp}=
0.732\,B_{12} / T_6 \gg 1$,
where the protons are strongly quantized.
The interaction of a magnetic field with proton spin,
according to \req{DeltaF}, yields
\beq
   { C_{Vp}^{(2)} } = \left[ {\beta g_p\hbar\omega_{cp}
         \over 4\cosh(\case14\beta g_p\hbar\omega_{cp}) }
            \right]^2\,N_0,
\eeq
which vanishes in the limiting cases of 
$\hbar\omega_{cp}\ll k_B T$ and $\hbar\omega_{cp}\gg k_B T$.
In the latter case, $C_V/N_{\rm tot}$ 
would tend to $\case12$ for the fully ionized plasma.
In Fig.~\ref{fig-cvm}, however, this does not happen
because of the contribution of neutral atoms,
which are subject to thermal ionization in this $\rho$-$T$-$B$ domain.
On the contrary, $C_V$ increases with increasing $B$,
since the neutral fraction becomes larger.
The two humps visible on each magnetic isotherm
correspond to the regions of the pressure destruction
of the first excited atomic state $s=1$, $\nu=0$
and the ground state $s=\nu=0$, respectively.
In the latter case, $C_V$ even exceeds the nonmagnetic
value, because of the delayed onset of degeneracy.
Only with density approaching $\rho_B$,
is the zero-field value of the heat capacity recovered.

This illustrates the 
main effects of a strongly quantizing magnetic field
on a partially ionized hydrogen plasma.
Other thermodynamic quantities,
obtained within the framework of the present model,
experience similar profound modifications.

\section{Conclusions}
We have developed a thermodynamic model of the 
hydrogen plasma in strong magnetic fields,
making use of the available quantum-mechanical results for
the fully ionized plasma and for the hydrogen bound
species. 
Applicability of the developed model is limited to
the temperatures $T$, densities $\rho$, and magnetic field 
strengths $B$ at which formation
of molecules and other bound species more complex than 
the H atoms may be neglected.
This condition holds, for instance, at $B_{12}\lesssim10$ and 
$T\gtrsim10^6$~K (any $\rho$)
or at $T\gtrsim10^5$~K and 
$\rho \lesssim 10^4\,(T_6 / B_{12})^3\gcc$.
Furthermore, although the theory presented 
in Sec.~\ref{sect-PI} is rather general,
our numerical results in partially ionized regions
are restricted to $B_{12}\geq0.7$, because
fitting formulas \cite{P98}
for quantum-mechanical characteristics of the
atoms moving in magnetic fields have been derived under this condition.
This restriction is fulfilled for a majority of neutron stars.
For laboratory field strengths (at $\gamma\ll1$),
perturbative methods may be sufficient.

Calculations in the frames of our model show that 
the magnetic field effects
strongly modify the thermodynamic functions 
and phase diagram of the plasma,
in particular the partial ionization region.
The abundance of atoms is significant
in the considered domain of temperatures
$T\sim10^5-10^{6.5}$~K and magnetic field strengths
$B\sim10^{12}-10^{13}$~G
at densities up to $\rho\sim10^2-10^3\gcc$,
contrary to the zero-field case.
At relatively low densities ($\rho\lesssim 1-100\gcc$,
depending on $B$ and $T$), the decentered atomic 
states possessing a large constant dipole moment
are significantly populated.
Since these values of $\rho$, $T$, and $B$ are typical of 
atmospheres of isolated neutron stars, 
the physical effects discussed above are expected 
to affect the spectra. 
It has been shown \cite{PP95,PP97} that
the presence of a nonionized component
and, in particular, decentered atoms
should produce observable absorption
and thus necessitate a modification
of previous fully ionized atmosphere models\cite{Shib92}.
Work in this direction is under way\cite{PSV}.

\begin{acknowledgements}
A.Y.P.\ is grateful to the theoretical astrophysics group
at the Ecole Normale Sup\'erieure de Lyon for hospitality 
and financial support.
A.Y.P.\ and Y.A.S.\ acknowledge useful discussions with J.~Ventura
and (at the initial stage of the work)
with G.~G.\ Pavlov and V.~E.\ Zavlin.
We thank M.~Steinberg for sending us
some of his results prior to publication.
The work of A.Y.P.\ and Y.A.S.\ has been partially
supported by INTAS Grant No.\ 96-542
and RFBR Grant No.\ 99-02-18099.
\end{acknowledgements}


\begin{references}
\bibitem{size}
Following\cite{P96}, we estimate $l_n=a_0 n\,\sqrt{7n^2+5}\,/2$
for an atom with the principal quantum number $n$.
This yields $d_{00}=a_0\,\sqrt{3}$, which is 13\%
smaller than the SC\cite{SC} hard-core diameters.
\bibitem{endnote1}
In the partially degenerate regime, 
convergence of this procedure
may be poor. In that case we impose 
a stabilization restriction that $w_\kappa$
does not vary appreciably from its value at the preceding iteration.
\end{references}
\end{document}